\documentclass[letter]{aa} 

\usepackage{graphicx}
\usepackage{txfonts}

\begin{document} 

\title{Temperature inversion in a gravitationally bound plasma: Case of the solar corona}
   \author{Luca Barbieri
          \inst{1,2,3}
          \and
          Lapo Casetti
          \inst{1,2,3}
          \and
        Andrea Verdini
          \inst{1,2}
        \and
        Simone Landi
            \inst{1,2}
          }
   \institute{Dipartimento di Fisica e Astronomia, Università di Firenze, via G.\ Sansone 1, I--50019 Sesto Fiorentino, Italy
     \and
     INAF -- Osservatorio Astrofisico di Arcetri, Largo Enrico Fermi, 5, I-50125 Firenze, Italy
     \and      
     INFN, Sezione di Firenze, via G. Sansone 1, I--50019 Sesto Fiorentino, Italy
             }
   \date{Received xxx, yyyy; accepted xxx, yyyy}
  \abstract 
  {The temperature of the solar atmosphere increases from thousands to millions of degrees moving from the lower layer (chromosphere) to the outermost one (corona), while the density drops accordingly. The mechanism behind this phenomenon, known as a temperature inversion, is still unknown. In this work, we model a coronal loop as a collisionless plasma confined in a semicircular tube that is subject to the Sun's gravity and in thermal contact with a fully collisional chromosphere behaving as a thermostat at the loop's feet. By using kinetic $N$-particle simulations and analytical calculations, we show that rapid, intermittent, and short-lived heating events in the chromosphere drive the coronal plasma towards a {non-equilibrium} stationary state. 
{The latter} is characterized by suprathermal tails in the particles' velocity distribution functions, exhibiting temperature and density profiles strikingly similar to those observed in the atmosphere of the Sun. These results suggest that a million-Kelvin solar corona can be produced without the local deposition of heat in the upper layer of the atmosphere that is  typically assumed by standard approaches. We find that suprathermal distribution functions in the corona are self-consistently produced instead of postulated a priori, in contrast to classical kinetic models based on a velocity filtration mechanism. }
 
   \keywords{Sun: corona -- Sun: atmosphere -- plasmas -- methods: numerical -- methods: analytical}
   \maketitle
   
\titlerunning{Temperature inversion in a gravitationally bound plasma}   
\authorrunning{L.\ Barbieri et al.}
\section{Introduction}
Temperature inversion, namely, an anticorrelation between temperature and density profiles, occurs in astrophysical systems, such as the filaments in molecular clouds \citep{Arzoumanian:aa2011,TociGalli:mnras2015}, the Io plasma torus around Jupiter \citep{MeyerVernetMoncuquetHoang:icarus1995}, the Earth's magnetosphere \citep{Ma:JGRSpacePhys2020}, the hot gas in some galaxy clusters \citep{BaldiEtAl:apj2007,WiseMcNamaraMurray:apj2004} and, notably, the solar atmosphere, which is with no doubt the most striking and thoroughly studied example (see e.g.,\ \citealt{GolubPasachoff:book} and references therein). 
The Sun's outermost layer, the corona, reaches temperatures above $10^6~\mathrm{K. It}$  lies on top of a denser and cooler (roughly $10^4~\mathrm{K}$) layer, the chromosphere, with the two layers connected by the so-called transition region, a thin interface only hundreds of kilometers wide that sees temperature variations up to a factor of 50 and density drops by the same factor \citep{observedtemperature}. The mechanism behind this "coronal heating problem"  is still largely unknown \citep{Klimchuk_2006,2012coronalheating}.

Assuming local thermodynamic equilibrium, a corona can form only upon the local deposition of heat in the upper layer. 
The energy coming from the Sun is enough to heat the whole corona \citep{Withbroe:1981tw}, bringing on the question of how energy is transported and dissipated at coronal heights. Some proposed mechanisms involve the release of magnetic energy stored in the corona \citep{Parker:1972wu,Dmitruk:1997uf,2005ApJ...618.1020G,Rappazzo:2008vl,2013ApJ...773L...2R,2015RSPTA.37340265W}, transmission and damping of waves generated by photospheric motions \citep{Heyvaerts_Priest_1983,Ionson_1978,2020A&A...636A..40H}, or a direct insertion of hot plasma in form of chromospheric spicules \citep{Pontieu:2011vg}.
However, there are theoretical indications \citep{1983ApJ...266..339S,2000ApJ...532L..71E} and observational evidence \citep{Dudik:2017un} to support that local thermodynamic equilibrium may not be satisfied in the transition region and in the corona.
This allows for another way of obtaining temperature inversion, as first recognized by \cite{Scudder1992a,Scudder1992b}. If the velocity distribution functions (VDFs) of electrons and ions have suprathermal tails\footnote{Kappa functions \citep{lazar2021kappa} were employed by \cite{Scudder1992a,Scudder1992b} to solve the problem analytically.} in the chromosphere, then temperature must increase with height: faster particles are able to climb higher in the Sun's gravity well and temperature increases thanks to ``velocity filtration.'' Notably, this implies no local deposition of heat in the upper layer.
Unfortunately, the model does not produce a transition region and, more importantly, non-thermal distributions in the strongly collisional chromospheric plasma are difficult to justify.

Interestingly, recent numerical studies have shown that when an isolated system governed by long-range interactions is impulsively perturbed it undergoes collisionless relaxation, ultimately reaching a non-thermal stationary state with temperature inversion, whose origins can be traced back to velocity filtration \citep{Casetti2014,Teles-Casetti2015,Gupta_2016}. 
This might explain the inverted temperature-density profiles of filaments in molecular clouds \citep{DiCintio2017}, but it cannot be applied as such to the coronal plasma, which is not isolated, but rather engaged in steady thermal contact with the chromosphere. 
However, while the VDFs of the chromospheric plasma are likely to be thermal due to collisionality, the chromosphere is a very dynamic environment, showing fine-scale structures down to instrumental resolution \citep{Cauzzi:2009ta,Ermolli:2022um}, while its temperature is expected to fluctuate in space and time \citep{Peter:2014uz,Hansteen:2014us}.

In this letter, using both numerical simulations and analytical modelling, we show that rapid temperature fluctuations in the chromosphere are able to build up a hot corona.
For simplicity, we modeled a coronal loop as a semicircular tube of collisionless plasma in thermal contact with a thermostat, the latter playing the role of the dynamic and collisional chromospheric plasma. 
The tunable part of the model is the statistics of the temperature fluctuations of the chromosphere. Temperature inversion appears as a very robust phenomenon and does not require any fine-tuning. However, matching the density and temperature values of the solar corona, while obtaining a (thick) transition region, requires a specific range of temperature fluctuations. Here, we briefly discuss whether the latter may be observationally tested.   

\section{Model and numerical simulations}
We consider a model of the plasma in a coronal loop made up of 
$N_e=N_i=2N$ protons and electrons having masses $m_i$ and $m_e$ and charge $e_i=-e_e=e$.  
Particles are confined in a semicircular tube of length, $2L,$ and cross-section, $S$, and they are subject to a constant downward gravity and to an electric force that ensures charge neutrality, 
whose combined effect is proportional to $(m_e+m_i)/2$ 
\citep{Pannekoek_1922,Rosseland_1924,belmont2013collisionless}. 
We assume that all quantities depend only on the coordinate along the loop axis, $x\in [-L,L]$ and are symmetric with respect to the loop top located at $x = 0$. We expand the electrostatic interactions between the particles in a Fourier series, with wave vectors $k_n=\pi n x/L$, only up to $n=1$. This is a mean-field approximation, where each particle only experiences the average effect of all the others, analogous to that yielding the Hamiltonian mean field model for single-component long-range-interacting particle systems \citep{AntoniRuffo:pre1995,Chavanis2005,GiachettiCasetti:jstat2019}. With these assumptions,   
the equations of motion are: 
\begin{equation} \label{HMF2Sne}
    m_\alpha \ddot{x}_{j,\alpha} =e_{\alpha}E(x_{j,\alpha})+g\frac{m_e+m_i}{2}\sin{\biggl(\frac{\pi x_{j,\alpha}}{2L}\biggl)}~, \\
\end{equation}
where $j = 1,\ldots,2N$ numbers the particles, $\alpha=e,i$ denotes the species, $g=G M_{\odot}/R_{\odot}^2$ is the Sun's gravity, and
\begin{equation} \label{eq:E(x)}
    E(x)=8\frac{|e_{\alpha}|}{S} N \left(q_i - q_e\right)\sin{\biggl(\frac{\pi x}{L}\biggl)}
\end{equation}
is the self-consistent electric field\footnote{We use electrostatic cgs units.}, with
\begin{equation} \label{eq:strat_param}
q_{\alpha}=\frac{1}{2N}\sum_{j=1}^{2N} \cos{\biggl(\frac{\pi x_{j,\alpha}}{L}\biggl)}~. \\
\end{equation}
In Eq.\ \ref{eq:strat_param}, $q_{\alpha}$ is the ``stratification parameter'' for each species, obtained by averaging the cosine of the particle angular position along the loop. As a result, $q_\alpha=0$ corresponds to a uniform distribution of particles, $q_\alpha = -1$ (resp.\ $+1$) to a distribution concentrated at the base of the loop, in $x = \pm L$ (resp.\ at the top, in $x = 0$). When $q_{\alpha} \in (-1,0)$ (resp.\ $(0,1)$) the density decreases (resp.\ increases) from the bottom to the top of the loop. The difference $q_i-q_e$ in Eq.\ \ref{eq:E(x)} measures the charge imbalance giving rise to the electric field.
The anchoring of the loop feet to the high chromosphere is modeled by coupling the coronal plasma to a thermostat at $x = \pm L$: 
when a particle reaches the boundary, it is reinjected into the loop with a random velocity consistent with the flux from a thermal distribution at the temperature of the thermostat \citep{Landi-Pantellini2001,LepriMPC2021,ThermalWalls}. 
Once written using dimensionless variables, the equations of motion depend only on three parameters: 
the mass ratio, $M={m_i}/{m_e}$, and the strengths of the interactions, $C$ and $\tilde{g}$, expressed as the electrostatic and gravitational energy, respectively, normalized to the thermal energy,
\begin{equation}
C  =  \frac{8n e^2 L^2}{\pi k_B T_0 } ~ \text{and} ~ \tilde{g}  =  \frac{g L (m_i+m_e)}{2 \pi k_B T_0}~, \label{eq:Cgtilde}
\end{equation}    
where $n$ is the the average number density of each species, $k_B$ is the Boltzmann constant, and $T_0 = 10^4$ K is the reference temperature of the thermostat and our unit of temperature.

Starting with the plasma in thermal equilibrium at the initial thermostat's temperature, $T = 1$ (in our units), the latter is allowed to vary so as to mimic the dynamic nature of the chromosphere.
After a waiting time, $t_w$, we increase the thermostat temperature by a quantity $\Delta T$, we keep the thermostat at $T = 1 + \Delta T$ for a time, $\tau$, and then we switch back to $T = 1$. These steps are iterated for the whole duration of the simulation run, drawing the values of $t_w$ and $\Delta T$ from two exponential distributions,
\begin{equation} \label{eq:gandf}
\eta (t_w)=\frac{1}{\langle t_w \rangle}e^{-\frac{t_w}{\langle t_w \rangle}}~\text{and}~\gamma(\Delta T)=\frac{1}{T_p}e^{-\frac{\Delta T}{T_p}}~.
\end{equation}
The physical idea is that the chromospheric temperature fluctuates due to random heating events, among which those producing a larger $\Delta T$ are less likely to occur. We observe that when both $t_w$ and $\tau$ are sufficiently small and, in particular, when these values are smaller than the relaxation times to equilibrium, the plasma in the loop is kept away from the initial thermal state, never relaxes to a thermal distribution with $T > 1$;  rather, it attains a non-equilibrium stationary state, with non-thermal distribution functions and inverted temperature-density profiles.

We solved Eqs.\ \ref{HMF2Sne}-\ref{eq:strat_param}, coupled to the fluctuating thermostat, using a fourth-order symplectic algorithm \citep{Candy1991} with a time step of $\delta t = 10^{-4}$. Due to the mean-field nature of the particles' equations of motion, the computational cost scales as $N$ (instead of $N^2$ as in direct $N$-body codes) and in the limit of large $N,$ the dynamics is collisionless and equivalent to a Vlasov one \citep{Campa2014}. 

\section{Simulations results}
We now focus on the case of a coronal loop with average number density, $n = 2.5 \times 10^9$ cm$^{-3}$, and half-length, $L = \pi\times 10^4$ km, so that $\tilde{g} = 16.64$. For the mass ratio we chose the realistic value  $M = 1836$. 
A realistic value of the parameter $C \approx 10^{22}$ implies very fast plasma oscillations and prohibitively small integration time steps ($\approx1/\sqrt{C}$) to reliably simulate the dynamics. However, we checked that in the stationary state the temperature and density profiles as well as the VDFs do not depend on the value of $C$, which only affects the amplitude and frequency of the oscillations around the means. 
Therefore, we arbitrarily chose $C = 400$ so that we may have oscillations with a periodicity greater than the thermostat timescale (see below), meaning that our simulated particles have a much smaller charge than in the real world. The simulations were performed using $N \approx 2.1 \times 10^6$, $\tau = 8\times 10^{-4}$, $\langle t_w\rangle = 4\times 10^{-2}$, and $T_p = 90$, i.e., $9\times 10^5$ K in physical units. Together with the choice of $T_0 = 10^4~\mathrm{K}$, these values imply an average temperature $T_b \approx 1.2 \times 10^4~\mathrm{K}$ at the base of the loop (see below). This sets the base at a height $z_b \approx 2\times 10^3~\mathrm{km}$ above the chromosphere, where the transition region begins, and the top of the loop at $z_t \approx 2.2\times 10^4~\mathrm{km}$, well into the corona. 

In the top panel of Fig.\ \ref{fig:QSS} we report the time evolution of the kinetic energies per particle
\begin{equation} \label{eq:K_alpha}
K_{\alpha} = \frac{1}{2N}\sum_{j=1}^{2N} \frac{p_{j,\alpha}^2}{2M_{\alpha}}~ ,     
\end{equation}
where $p_{j,\alpha}$ is the dimensionless momentum of the $j$-th particle of species $\alpha$, with $M_{e} = 1$ and $M_{i} = M$; in the bottom panel the stratification parameters $q_e$ and $q_i$ defined in Eq.\ \ref{eq:strat_param} are reported, again as a function of time. After a transient, a stationary state is reached where all these quantities fluctuate around constant values, the transient being longer and the amplitude of the fluctuations smaller for the ions, as expected due to their larger mass. The initial value of $q_i=q_e \approx -0.99$ corresponds to the stratification of an isothermal atmosphere at $T=1$. In the stationary state, which is not a thermal equilibrium state (as we show in the following), a smaller value is obtained, $q_i\approx q_e\approx -0.94$, indicating a slightly milder stratification.

In Fig.~\ref{fig:temperatureinversion}, we report the temperature and density profiles of electrons along the loop, obtained by time-averaging in the stationary state\footnote{Temperature and density profiles of ion and electrons are equal in the stationary state since they experience the same temperature fluctuations at the chromosphere and feel the same external field.}. To make the comparison with the known profiles of the solar atmosphere easier, we report the results in physical units as a function of the height, $z,$ above the chromosphere (temperatures in K, densities in cm$^{-3}$, height in km). A temperature inversion is clearly visible, as well as the presence of a transition region: indeed, there is a sharp increase (resp.\ decrease) of the temperature (resp.\ density) roughly between $z = 2\times 10^3$ and $z = 5\times 10^3~\mathrm{km}$, with a jump from $1.2\times 10^4$ to $6\times 10^5~\mathrm{K}$ while density drops by two orders of magnitude, followed by a gentler change of both quantities. Both the values of $T$ and $n$ and the shape of the curves are strikingly similar to those of the Sun atmosphere (see e.g.,\ \citealp{GolubPasachoff:book,observedtemperature}), aside from the fact that the transition region in our model is larger than in the real case, that is, roughly $3\times 10^3~\mathrm{km}$ against less than $2\times10^2~\mathrm{km}$. 
\section{Theory}
\begin{figure}
    \centering
    \includegraphics[width=0.99\columnwidth]{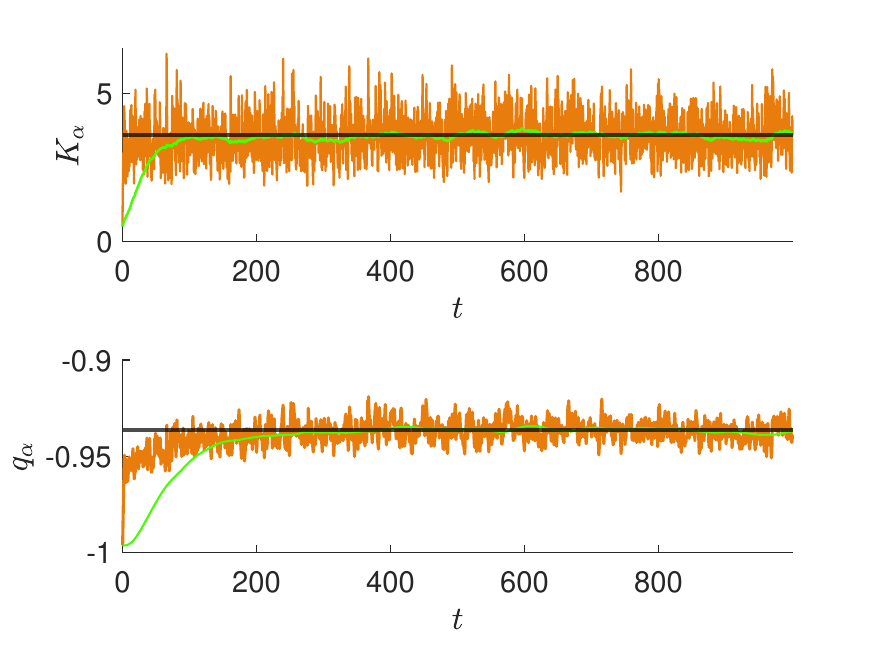}
    \caption{Kinetic energies (top) and stratification parameters (bottom) of ions and electrons in green and orange, respectively, as a function of time. The black horizontal lines are the theoretical stationary values predicted using Eq.\ \ref{eq:falphastationary}.
    }
    \label{fig:QSS}
\end{figure}

\begin{figure}
    \centering
    \includegraphics[width=0.99\columnwidth]{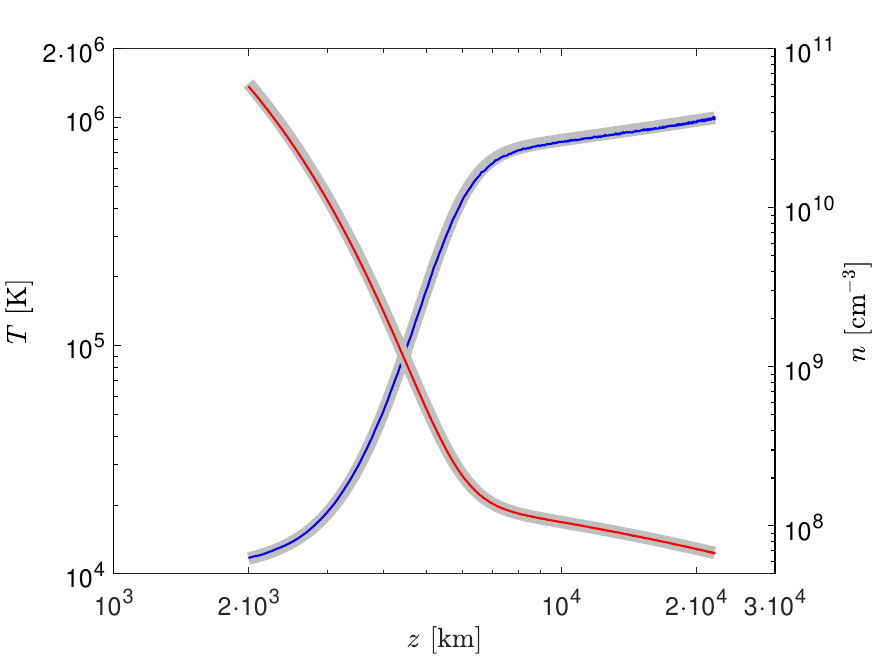}
    \caption{Electron density (red) and temperature (blue) as a function of the height above the photosphere, obtained in a numerical simulation with parameters corresponding to the solar atmosphere (see text). The thick grey lines are the theoretical profiles computed using Eq.\ \ref{eq:falphastationary}. Ion profiles (not shown) are the same.}
    \label{fig:temperatureinversion}
\end{figure}

\begin{figure}
    \centering
    \includegraphics[width=0.99\columnwidth]{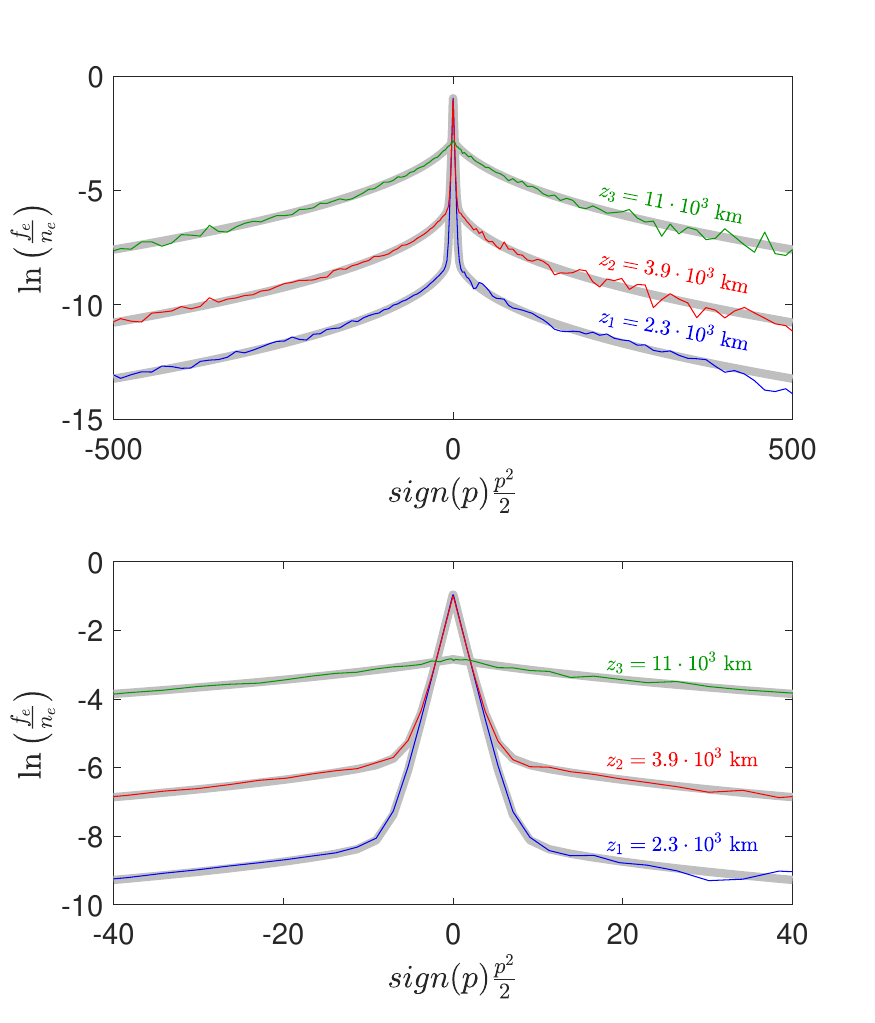}
    \caption{Electron VDFs (color curves) normalized by the electron densities as a function of the signed kinetic energy, at different heights (see labels). 
    Theoretical VDFs computed from Eq.\ \ref{eq:falphastationary} are plotted as thick grey lines. Ion VDFs (not shown) are the same if plotted against $\text{sign}(p)p^2/2M$.
    In the bottom panel, a magnification of the central region of the same VDFs is shown to highlight the disappearance of the Gaussian profile.}  
    \label{fig:VDFe}
\end{figure}

The numerical results can be reproduced and explained computing the distribution functions (DFs) $f_\alpha(\vartheta,p)$ of ions and electrons in the non-equilibrium stationary state ($\vartheta$ is the dimensionless position). To this end, we first consider the time averaged incoming flux of energy at the feet of the loop, i.e., at $\vartheta = \pm \pi$, which can be written for each species as $J_{IN,\alpha} = \langle J_{T(t),\alpha} \rangle_t$, where we have
\begin{equation}
J_{\zeta,\alpha} = \int_0^{+\infty}dp \, \frac{p}{\zeta M_{\alpha}} \, e^{-\frac{p^2}{2 \zeta M_{\alpha}}} 
,\end{equation}
which is the flux when the thermostat is at temperature $\zeta$. 
To explicitly compute $J_{IN,\alpha}$ we replace time averages with averages over the probability distributions of waiting times and of thermostat temperatures, obtaining
\begin{equation} \label{eq:averagefluxin}
    J_{IN,\alpha} = (1-A) J_{1,\alpha} + A\langle J_{1+\Delta T,\alpha} \rangle_\gamma  \, ,
\end{equation}
where $A=\tau/\left(\tau  + \langle t_w \rangle_\eta \right)$ and $\langle \cdot \rangle_\gamma$ and $\langle \cdot \rangle_\eta$ stand for the averages over the distributions $\gamma$ and $\eta$ that are given in Eq.~\ref{eq:gandf}. In our model the plasma in the loop is collisionless, so that $f_\alpha$ obeys the Vlasov equation and in the stationary state it depends on $\vartheta$ and $p$ only via the single-particle Hamiltonian,
\begin{equation} \label{eq:singleparticleH}
    H_{\alpha}(\vartheta,p) =\frac{p^2}{2M_{\alpha}}+2\tilde{g}\cos{\left(\frac{\vartheta}{2}\right)} ~,
\end{equation}
as implied by the Jeans theorem. We also note that  an electrostatic interaction term proportional to the effective electric field should be present in Eq.~\ref{eq:singleparticleH}, but it can be proven that it vanishes in the stationary state \citep{HMF2Sextended}.  The DF at $\vartheta = \pm \pi$ is fixed imposing equality of the incoming and outgoing fluxes at the feet of the loop, that is, $J_{IN,\alpha} = J_{OUT,\alpha}$, with 
\begin{equation}\label{eq:jinjout}
J_{OUT,\alpha} = \int_{-\infty}^{0}dp\,  p \, f_{\alpha}(\pm \pi,p)\, .
\end{equation}
The DFs $f_\alpha(\vartheta,p)$ in the stationary state can be obtained using the distribution at the boundary combined with the Jeans theorem \citep{HMF2Sextended} because the level sets of the single-particle Hamiltonian \ref{eq:singleparticleH} are always connected to the bottom boundary (i.e., the thermostat), expressed as
\begin{equation} \label{eq:falphastationary}
    f_{\alpha}(\vartheta,p) = \mathcal{N}_{\alpha} \left[(1-A)\frac{e^{-H_{\alpha} (\vartheta,p)}}{M_{\alpha}} 
     + A\int_0^{+\infty} d\zeta \, \gamma(\zeta)\, \frac{e^{-\frac{H_{\alpha} (\vartheta,p)}{1 + \zeta}}}{(1+ \zeta)M_{\alpha}} \right]~,
\end{equation}
where $\mathcal{N}_{\alpha}$ values are such that $\int_{-\infty}^{+\infty} dp \int_{-\pi}^{\pi} d\vartheta \, f_{\alpha}(\vartheta,p) = 1$. 
The interpretation of Eq.~\ref{eq:falphastationary} is as follows: the stationary DF is given by a thermal distribution at temperature $T = 1,$ plus a non-thermal contribution arising from the average of thermal distributions at $T = 1 + \zeta$ over the probability distribution $\gamma(\zeta)$ of the temperature fluctuations. The weight of the non-thermal contribution is proportional to $A$, the fraction of time in which the thermostat is not at temperature $T = 1$. The thermal population dominates at small heights $z$, and is depressed by the gravity term in $H_\alpha$ when increasing $z$; conversely, the non-thermal contribution becomes more and more relevant at larger $z$ due to velocity filtration, because faster particles can climb higher in the potential well, showing up as suprathermal tails in the distribution. 
This is shown in Fig.~\ref{fig:VDFe}, where the VDFs of electrons normalized by the density are plotted at three increasing heights, $z=2.3,~3.9,~11\times10^3~\mathrm{km}$ from bottom to top, corresponding to the base of the transition region, the middle transition region, and the corona, respectively. Red, blue, and green curves are obtained from the simulation, while grey curves are the theoretical predictions of Eq.~\ref{eq:falphastationary}.
We note that VDFs are plotted as functions of the signed kinetic energy $\text{sign}(p)\, p^2/2$ in a semilogarithmic scale, so that a thermal distribution (a Gaussian) appears as a triangle symmetric about zero.  
The VDFs are always composed of a thermal core at small velocities plus a suprathermal tail at larger velocities. As the height increases, the thermal core progressively shrinks and almost disappears in the corona, where the VDF is basically suprathermal. 
These VDFs explain the shape of the temperature and density profiles reported in Fig.\ \ref{fig:temperatureinversion}, where the grey curves are the theoretical prediction obtained with Eq.~\ref{eq:falphastationary}. 
The sharp temperature rise in the transition region (and density decrease) is caused by the dramatic change in the shape of the VDF that passes from being nearly thermal to almost completely suprathermal.
Once coronal heights are reached, totally non-thermal VDFs produce a gentler variation of temperature and density, similar to the case described in \cite{Scudder1992a,Scudder1992b}.
For completeness, the stationary values of kinetic energies and stratification parameters computed using Eq.~\ref{eq:falphastationary} are drawn with black horizontal lines in Fig.~\ref{fig:QSS}. 

\section{Discussion and conclusions}
 We present a kinetic model of the solar atmosphere, where the collisonless coronal plasma is in steady contact with a thermostat mimicking a completely collisional chromosphere. The analytical and numerical results consistently show that in response to 
 intermittent rapid and short-lived increments of the chromospheric temperature, suprathermal tails in the VDFs naturally form and gravitational filtering causes a sharp temperature rise and density decrease in the above atmosphere, consisting of a (thicker than observed) transition region, followed by an extended corona at roughly $10^6$ K.

Suprathermal electrons, along with a thermal population, are measured in situ in the solar wind \citep{Pilipp_al_1987a,Halekas_al_2020,Maksimovic_al_2020}.
Their presence in the transition region \citep{Dudik:2017un} or in flaring regions \citep{Polito:2018ut} is also compatible with remote sensing observations of non-thermal line widths, but their direct detection is still challenging.
Our model supports the formation of a nonthermal population already at the base of the corona for closed loop geometry. We expect our mechanism to be valid also in the case of open field lines, thus favouring the formation of the solar wind by gravitational filtering as in exospheric models \citep{Jockers1970,Lemaire1971,Lamy2003,Maksimovic1997,Zouganelis2004}. However, in an open spherical geometry, there are also escaping particles, forming the wind, and ``trapped particles'' (whose trajectories never reach the base); thus, a more complex formalism would be necessary to tackle this case \citep{Zouganelis2004}. 

One aspect that is fundamental in achieving temperature inversion is that the coronal plasma is in a non-equilibrium stationary state.
This requires the chromosphere to maintain any given temperature for intervals much shorter than the relaxation time of electrons in the corona, $t_R$, which in our model, is expected to be the shortest between the sound travel time and the free fall time, that is, roughly $10$~s for the parameters we considered above.
To reach temperatures around $10^6$~K in the corona, while keeping temperatures around $10^4$~K at the base of the transition region, the mean of the chromospheric temperature increments must be as large as the coronal temperature, namely, $T_p \approx 10^6$~K, but the ratio between their duration, $\tau,$ and the typical waiting time, $\langle t_w \rangle,$ between them must be small, so that $\tau \ll \langle t_w \rangle \ll t_R$. 

Short-lived, intense, and small-scale brightenings are routinely observed on the Sun \citep{Dere:1989ux,Teriaca:2004wy,Peter:2014uz,Tiwari:2019us,Berghmans:2021wl}.
{Among the latter, the} so-called campfires, recently observed in extreme UV imaging by Solar Orbiter, have temperatures of $\approx 10^6$ K, while explosive events appearing in $\mathrm{H}\alpha$ line widths have smaller temperatures, around $2 \times 10^5$ K \citep{Teriaca:2004wy}, but are ten times more frequent. 
This trend is compatible with the exponential distribution of temperature increments we assumed. However, current measurements have a temporal resolution which is at best of a few seconds, so that temperature increments able to form a corona according to our model remain unresolved. Anyhow, even in the case of unresolved events, a given temperature measurement brings a footprint of the underlying event distribution. For instance, assuming $\langle t_w \rangle \approx 1$~s, with $\tau \approx 0.1$ s our model would predict an average temperature $T \approx 2 \times 10^4$~K at the top of the chromosphere, which is not observed; however, if $\tau \approx 1/50$~s we get $T \approx 1.2 \times 10^4$ K, as in the case we have shown in Fig.\ \ref{fig:temperatureinversion}. This is consistent with ALMA observations of brightness temperatures as large as $1.2 \times 10^4$~K \citep{Molnar_2019}, with measurements at a 2 s cadence. 
A possible physical mechanism at work is magnetic reconnection in the low atmosphere, as suggested by recent observations \citep{Lee:ApJ2020,Rauoafi:ApJ2023}.

The main result in this letter is that the average temperature profile of the solar atmosphere can be obtained keeping the coronal plasma in thermal contact with the chromosphere, without any local deposition of heat in the corona. 
At variance with previous results based on velocity filtration, a transition region appears, and non-thermal chromospheric VDFs are not needed: a corona can be formed out of a collisional, albeit dynamic, chromosphere. Clearly, this does not exclude that also other mechanisms contribute to coronal heating.

Potentially important physics is neglected in the model presented here, such as radiation losses, collisions in the corona, and magnetic fields. Although a detailed discussion of these effects is beyond the scope of this work, we address them in brief as follows. In our model, energy is transported in the corona by the particles' stream, and is exchanged with the thermostat on a timescale of the order of the crossing time of the loop, which is less than a minute and much smaller than the radiation timescale (around 30 m, see e.g.,\ \cite{Serio:AandA1991}, \cite{ShibataMagara:2011} and \cite{AntolinFroment:2022}), so that energy losses due to radiation can be safely neglected in our model. Moreover, radiation losses would not change the picture, because in the non-equilibrium state, a heat flux towards the radiating region would set in and restore the temperature profile \citep{1983ApJ...266..339S,Landi-Pantellini2001,Dorelli_Scudder_1999}. Collisions between particles will surely occur in the corona. In any case, thanks to the fact that the mean free path of a particle in a plasma scales with velocity $v$ as $v^4$, we expect that only ``cold'' particles would be heavily affected by collisions; specifically, the VDFs would become closer to thermal at small energies, while non-thermal features related to the ``hot'' particles, which are the ones selected by gravitational filtering and are then able to reach coronal heights, would not be erased by collisions \citep{1983ApJ...266..339S,Landi-Pantellini2001}. Velocity filtration is then expected to become more efficient in presence of some degree of collisionality; the latter effect, together with particle confinement by magnetic lines, may result in a transition region sharper than the one obtained here and closer to the observed one.

The mechanism that ends up producing a temperature inversion in the solar corona in our model is  general and it could prove relevant to other systems where hot coronae are thought to be present: stars other than the Sun, as well as active galactic nuclei and so on. 
\begin{acknowledgements}
We thank G.\ Cauzzi, P.\ Di Cintio, and E.\ Papini for very useful discussions and the anonymous referee for comments which helped us to improve the presentation of our work. We acknowledge partial financial support from the MIUR-PRIN2017 project \textit{Coarse-grained description for non-equilibrium systems and transport phenomena (CO-NEST)} n.\ 201798CZL, the National Recovery and Resilience Plan, Mission 4 Component 2 - Investment 1.4 - NATIONAL CENTER FOR HPC, BIG DATA AND QUANTUM COMPUTING - funded by the European Union - NextGenerationEU - CUP B83C22002830001, the Solar Orbiter/Metis program supported by the Italian Space Agency (ASI) under the contracts to the National Institute of Astrophysics (INAF), Agreement ASI-INAF N.2018-30-HH.0, and the Fondazione CR Firenze under the projects \textit{HIPERCRHEL} and \textit{THE SWITCH}. 
\end{acknowledgements}
   \bibliographystyle{aa} 
   \bibliography{manuscript} 
\end{document}